\documentstyle[12pt,epsf]{article}
\newcommand{\bm}[1]{\mbox{\boldmath $#1$}}
\def\e{\mbox{e}}
\def\eps{\epsilon}
\def\be{\begin{equation}}
\def\ee{\end{equation}}
\def\d{\partial}
\def\bk{b_{\mbox\scriptsize{\bf k}}}
\begin{document}

\begin{titlepage}
\title{$O(4)$ symmetric singular solutions and multiparticle cross
sections in $\phi^4$ theory at tree level}
\author{ F.L. Bezrukov, M.V. Libanov and S.V. Troitsky\\
{\small{\em Institute for Nuclear Research of the Russian Academy of
Sciences, }}\\
{\small{\em 60th October Anniversary prospect, 7a, Moscow
117312}}\\ }
\date{July 31, 1995}
\end{titlepage}
\maketitle

\begin{center}
{\bf Abstract}
\end{center}
We solve the classical euclidean
boundary value problem for tree-level multiparticle
production in $\phi^4$ theory at arbitrary energies in the
case of $O(4)$ symmetric
field configurations.  We reproduce known low-energy results and obtain
a lower bound on the tree cross sections at arbitrary energies.
\newpage

{\bf 1.}
In recent years,
remarkable progress
has been achieved
in calculating the amplitudes of multiparticle production in scalar
theories like ${\lambda\over 4}\phi^4$ (for a review see
ref.\cite{Voloshin-Rochester}). At the tree level, the amplitude of the process
$1 \to n$ has been calculated exactly at $n$-particle threshold
\cite{Voloshin-tree},
$$
  A_{1\to n}=n!\left({\lambda\over 8}\right)^{n-1\over2}
$$
It has been understood \cite{Brown} that this amplitude is related to
spatially homogeneous euclidean solution to the field equations which is
singular at $\tau=0$ ($\tau$ is euclidean time). The fact that singular
euclidean solutions play a role in calculating multiparticle amplitudes is not
too surprising: similar singular solutions are known to be relevant for
calculating semiclassical matrix elements in quantum mechanics by the Landau
technique \cite{Landau}. Furthermore, perturbative calculations about
the Brown solution \cite{LRST,LST} strongly suggest that to all orders
of the perturbation theory the multiparticle cross section in the limit
\begin{equation}
\lambda\to 0,~~~\lambda n=\mbox{fixed},~~~
\epsilon=(E-nm)/n=\mbox{fixed}
\label{1*}
\end{equation}
($E$ is the total center--of--mass energy) has the exponential form,
$$
\sigma_{1\to n}(n, E)\sim \exp\left[{1\over\lambda}F(\lambda n,\epsilon)\right]
$$
and that the leading exponent is independent of the initial state
provided the latter contains $O(1)$ particles. These features are again similar
to the semiclassical matrix elements in quantum mechanics calculable by the
Landau technique.

There have been several attempts to apply the Landau technique for calculating
multiparticle cross sections in field theories
\cite{Voloshin-1,Khlebnikov,DiakPetr,VolGor,Son}. In particular, Son \cite{Son}
formulated an appropriate classical boundary value problem for evaluation
of the
exponent $F(\lambda n, \epsilon)$ to all loops. Generaly, this approach
requires contours in comlex time plane, but at small $\lambda n$, the leading
term in $F$ can be calculated by studying purely euclidean s
ingular solutions
 of the field equation. In ordinary perturbation theory this leading term comes
 from tree graphs, so the dependence of $F$ on $\lambda$ is known explicitly,
\[F_{{\rm tree}}(\lambda n,E)=\lambda n \ln(\frac{\lambda n}{16})- \lambda
n +\lambda nf(\epsilon)\]
and the euclidean technique of Son \cite{Son} enables one, at least in
principle, to calculate the only unknown function, $f(\epsilon)$, of the
average kinetic energy of the outgoing particles, $\epsilon$. The loop
corrections add terms of order $(\lambda n)^2$ or higher into $F(\lambda n,
\epsilon)$.

Even in the simplest case of small $\lambda n$, the calculation of the exponent
$F_{{\rm tree}}$ at all energies $\epsilon$ is a complicated problem. The
corresponding solution has singularities on a three-dimentional hypersurface in
four-dimentional euclidean space (in this paper we consider four-dimentional
theories) \cite{Son}. This hypersurface depends on $\epsilon$ and has to be
 found in the process of calculation. In this paper we apply the Rayleigh-Ritz
procedure and consider $O(4)$ symmetric field configurations. Since
$F_{{\rm tree}}$ can be obtained by maximization over all hypersurfaces
\cite{Son}, our estimate of the exponent for the tree cross section is in
fact the lower bound on $F_{{\rm tree}}$ at given $\epsilon$.

{\bf 2.}
Let us consider the tree cross section of producing $n$ scalar particles
with total energy $E$ by a few virtual ones in a model with the lagrangian
\[L=\frac{1}{2}(\partial_\mu\phi)^2-\frac{1}{2}\phi^2-\frac{\lambda}{4}
\phi^4\]
where the mass of the boson is set equal to 1. As outlined above,
in the limit
(\ref{1*}) the tree cross section, with exponential accuracy, does not depend
 on the initial state and has the form
\[\sigma_{{\rm tree}}(E,n)\sim\exp\left[W_{{\rm tree}}(E,n)\right]\]
where
\begin{equation}
W_{{\rm tree}}(E,n)=\frac{1}{\lambda}F_{{\rm tree}}(\lambda n,\epsilon)=
n\ln\frac{\lambda n}{16}-n+nf(\epsilon)
\label{5**}
\end{equation}
The prescription of ref.\cite{Son} for calculating $W_{{\rm tree}}
(E,n)$
is as follows. One first introduces two extra parametrs, $T$ and $\Theta$,
which are Legendre conjugate to $E$ and $n$. Then one considers
hypersurfaces in euclidean space,
\begin{equation}
\tau=\tau(\bm{x})
\label{5*}
\end{equation}
with the condition that
\begin{equation}
\tau(\bm{x}=0)=0
\label{5+}
\end{equation}
The field configurations of interest are the solutions to the euclidean field
equations which are singular at the surfaces (\ref{5*}) and vanich at $\tau\to
+\infty$. For a given surface of singularities the solution to this boundary
value problem is argued to be unique; its spatial Fourier components
exponentially decay at $\tau\to\infty$,
\[
\phi({\bf k}, \tau)=\frac{1}{\sqrt{2\omega_{\bf k}}}\bk\exp
\left(-\omega_{\bf k}\tau\right)
\]
where $\omega_{\bf k}=\sqrt{{\bf k}^2+1}$. For a given surface of singularities
one defines the object
\begin{equation}
S(T,\Theta)=\int d^3k\bk^*\bk\exp\left(\omega_{\bf k}T-\Theta\right)
\label{6+}
\end{equation}
and its Legendre transform with respect to $T$ and $(-\Theta)$,
\begin{equation}
W(E,n)=ET-n\Theta-S(T,\Theta)
\label{6*}
\end{equation}
\[
{\d S\over \d T}=-E
\]
\be
{\d S \over \d\Theta}=n
\label{6++}
\ee
Note that Eq.(\ref{6++}) can be written as
\be
S=-n
\label{6**}
\ee
The function $W(E,n)$ still depends on the form of the singularity surface.
To obtain $W_{\rm tree}(E,n)$, one should maximize
(\ref{6*}) over all singularity surfaces subject to the condition (\ref{5+}).
Note that the specific dependence of $S(T,\Theta)$ on $\Theta$,
$$
S(T,\Theta)\propto \e^{-\Theta}
$$
ensures that $W(E,n)$ (and, correspondingly, $W_{\rm tree}(E,n)$) is
indeed parametrized by the only function $f(\epsilon)$, as indicated in
Eq.(\ref{5**}).

This program has been carried out analytically for small $\eps$
\cite{Son} and earlier perturbative results
\cite{LRST} have been reproduced,
\be
f(\eps)={3\over 2}\log{\eps\over 3\pi}+{3\over 2} - {17\over 12}\eps+
O(\eps^2)
\label{pert}
\ee
In the opposite limit, $\eps\gg 1$, the mass term in the field equation
can be neglected and the exponent of the cross section approaches the
limit $F(\lambda n,\infty)$
determined by the corresponding solution of the massless field
equations. Provided the correct saddle point is the maximum
over the singularity curves, one can
obtain a lower bound on the tree cross section in the ultra-high energy
regime from known \cite{Khlebnikov} massless singular
$O(4)$-symmetric solution,
\be
  \phi(\tau,\bm{x})=\sqrt{8\over\lambda}\,{\rho\over
  \bm{x}^2+(\tau+\rho)^2-\rho^2}
\label{3*}
\ee
($\rho$ is the radius of the 3-dimensional sphere where the solution is
singular), from which one obtains
$$
f(\infty)\ge\log(2/\pi^2)
$$

Entirely different way of obtaining lower bound on the tree cross section is
to estimate tree diagrams directly \cite{Voloshin}. We will compare our
results with this bound in what follows.

{\bf 3.} To estimate $f(\eps)$ entering Eq.(\ref{5**}) for all
$0\le\eps<\infty$
we apply the Rayleigh-Ritz procedure and consider $O(4)$ symmetric fields.
This will actually provide a lower bound for $\sigma_{\rm tree}(E,n)$ at
given $E$ and $n$. The singularity surfaces are then three-spheres in
euclidean
space, and the only variational parameter is the radius of a sphere, $\rho$.
Since the singularity surface should touch the origin (see Eq.(\ref{5+})),
the singularity sphere should be centered at
$$x_0=(-\rho,{\bf 0})$$
i.e., the $O(4)$ symmetric solution has the general form
$$
\phi=\phi(r)
$$
$$
r=\sqrt{\bm{x}^2+(\tau+\rho)^2}
$$
At large $r$, the solution tends to the exponentially falling solution to
the free field equation,
$$
\phi\propto{K_1(r)\over r}\propto{\e^{-r}\over r^{3/2}}
$$
i.e., at $\tau\to\infty$ one has
$$
\phi=A(\rho){\exp(-\sqrt{\bm{x}^2+(\tau+\rho)^2}) \over
(\bm{x}^2+(\tau+\rho)^2)^{3/4} }
$$
where the coefficient function $A(\rho)$ is to be determined by solving
the field equations
under the condition that it has a
singularity at $r=\rho$. From
this asymptotics one finds
$$
\bk=2A(\rho)
{\e^{-\omega_{\bf k}\rho} \over \sqrt{2\omega_{\bf k}} }
$$
The ''action'' (\ref{6+}) is then expressed through $A(\rho)$, so
$W$ (Eq.(\ref{6*})) reads
\be
W=ET-n\Theta-8\pi A^2(\rho)\e^{-\Theta} {K_1(2\rho-T) \over 2\rho-T}
\label{W}
\ee
Now we extremize Eq.(\ref{W}) over $\rho$, $T$ and $\Theta$.
This leads
to the following equations which determine the saddle point values of these
three parameters,
\be
{E\over n}={A'(\rho)\over A(\rho)}= {K_2(2\rho-T) \over K_1(2\rho-T)}
\label{7}
\ee
\be
\Theta=\ln \left\{
{8\pi A^2(\rho) \over n} \, {K_1(2\rho-T) \over 2\rho-T} \right\}
\label{5***}
\ee
  So, we look for the classical solutions which are singular at the
spheres $r^2=\rho^2$ and from their asymptotics obtain
$A(\rho)$, then express saddle point values of $\rho$, $T$ and $\Theta$
through $E$ and $n$ (by making use of the Eqs. (\ref{7}), (\ref{5***}))
and finally obtain the estimate for the exponent for the
tree cross section (see Eqs.(\ref{6*}) and (\ref{6**}))
$$
W_{\rm tree}=ET-n\Theta-n
$$
It is straightforward to perform this calculation numerically for all
$\epsilon$.

{\bf 4.} The result of numerical solution of the field equation with
singularities at different $\rho$, the function $A(\rho)$, is presented in
Fig.1. The resulting radius of the singularity sphere, $\rho(\eps)$,
is shown in Fig.2.
The exponent for the cross section indeed has the
form of Eq.(\ref{5**}) with the function $f(\eps)$ plotted in Fig.3.

At low energies our result
matches the perturbative results \cite{LRST}, Eq.(\ref{pert}).
The fact that our variational approach leads to the exact results for
$W_{\rm tree}$ at small $\eps$ can be understood as follows. At small
$\eps$, the curvature of the singularity surface is always large, and
only this curvature is relevant for the evaluation of $f(\eps)$
\cite{Son}. In other words, the surface of singularities has the form
$$
\tau(\bm{x})=\alpha\bm{x}^2+O(\bm{x}^4)+\cdots
$$
and only the leading term is important at small $\eps$. Clearly, this
leading term can be reproduced exactly in our $O(4)$ symmetric ansatz,
and our result is exact at small $\eps$.

At very high energies (small
values of $\rho$) our field configuration tends to the solution of the
massless equations, Eq.(\ref{3*}). So, at $\eps\gg 1$ it approaches the
lower bound derived from this solution.

The alternative lower bound on $f(\eps)$
can be easily read out from ref. \cite{Voloshin} and it
is also shown in Fig.3.
This bound has been obtained by direct analysis of
diagrams. As one can see from Fig.3, our new bound is stronger than that
of ref. \cite{Voloshin}.

   To summarize, we have solved numerically the boundary value problem for
tree
multiparticle cross-sections at arbitrary energy for spherically symmetric
field configurations. At low energies our solution reproduces the exact tree
results, while for general energy it gives the lower bound on the tree
cross section. To improve this estimate, one should consider more general
field configurations whose symmetry is at most $O(3)$.

We would like to thank A.N.~Kuznetsov, D.T.~Son, P.G.~Tinyakov
and especially V.A.Rubakov for
numerous helpful discussions. This work is supported in part by the ISF
grant \# MKT 300 and INTAS grant \# INTAS-93-1630.

\newpage
\pagestyle{empty}
\epsfbox{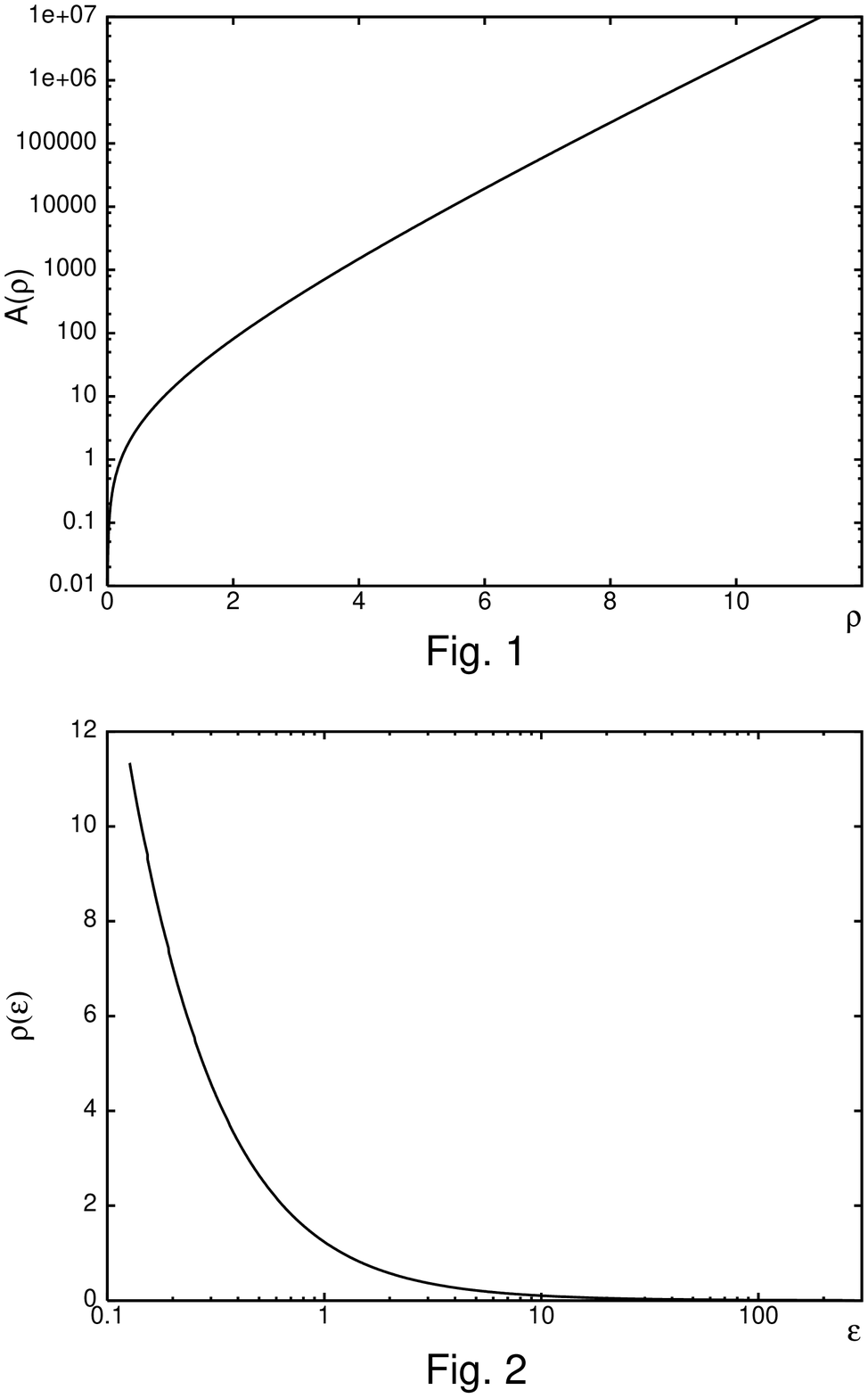}
\newpage
\epsfbox{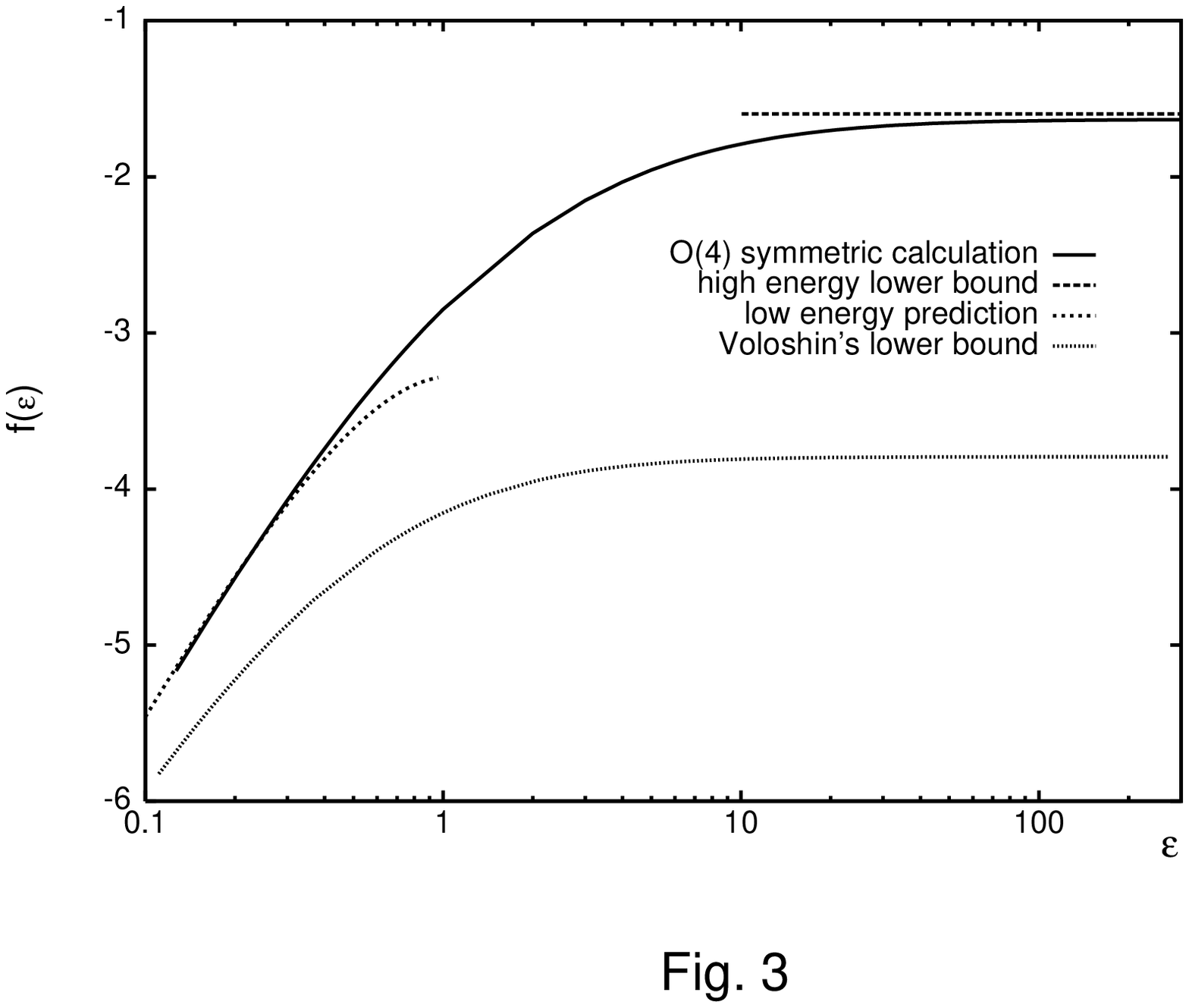}

\end{document}